\documentclass[aps,prl,showpacs,superscriptaddress,amsmath,amssymb,floatfix,twocolumn]{revtex4}
\usepackage{graphicx}
\usepackage{hyperref}
\usepackage{wasysym}

\begin{document}
\title{Spatially probed electron-electron scattering in a two-dimensional electron gas}

\author{M. P. Jura$^\ast$} \affiliation{Department of Applied Physics, Stanford University,
Stanford, California 94305, USA}
\author{M. Grobis$^\dag$} \affiliation{Department of Physics, Stanford University, Stanford, California 94305, USA}
\author{M. A. Topinka$^\ddag$} \affiliation{Department of Physics, Stanford University, Stanford, California 94305, USA} \affiliation{Department of Materials Science \& Engineering, Stanford University, Stanford, California 94305, USA}
\author{L. N. Pfeiffer$^\S$} \affiliation{Bell Labs, Alcatel-Lucent, Murray Hill, New Jersey 08544, USA}
\author{K. W. West$^\S$} \affiliation{Bell Labs, Alcatel-Lucent, Murray Hill, New Jersey 08544, USA}
\author {D. Goldhaber-Gordon$^\parallel$} \affiliation{Department of Physics, Stanford University, Stanford, California 94305, USA}


\begin{abstract}
Using scanning gate microscopy (SGM), we probe the scattering between a beam of electrons and a two-dimensional electron gas (2DEG) as a function of the beam's injection energy, and distance from the injection point.  At low injection energies, we find electrons in the beam scatter by small-angles, as has been previously observed.  At high injection energies, we find a surprising result: placing the SGM tip where it back-scatters electrons \textit{increases} the differential conductance through the system.  This effect is explained by a non-equilibrium distribution of electrons in a localized region of 2DEG near the injection point.  Our data indicate that the spatial extent of this highly non-equilibrium distribution is within $\sim 1 \ \mathrm{\mu m}$ of the injection point.  We approximate the non-equilibrium region as having an effective temperature that depends linearly upon injection energy.
\end{abstract}


\pacs{73.40.-c, 07.79.-v, 73.23.Ad}

\maketitle

\section{Introduction}

Electron-electron (e-e) interactions play a fundamental role in the behavior of electronic systems at low temperatures: leading to the formation of exotic quantum states \cite{Willett-Wigner} or, in Fermi liquids, determining the lifetime of quasiparticle excitations \cite{Giuliani}.  In many mesoscopic systems, including GaAs-based two-dimensional electron gases (2DEGs) which we study here, e-e scattering is the dominant source of inelastic scattering and hence dephasing \cite{Giuliani,AAK,Yacoby-DoubleSlit} for low energy electrons.  Nanoconstrictions allow the flow of non-equilibrium currents between two reservoirs, and e-e scattering therefore dictates energy exchange in and around nanoconstrictions.  E-e scattering has been observed to cause electron heating in metallic nanowires \cite{Devoret-wires} and hydrodynamic flow effects in 2DEG-based wires \cite{Hydrodynamic}.  Non-equilibrium currents through short constrictions, quantum point contacts (QPCs), in 2DEGs have been used as charge-detectors for quantum information processing \cite{Petta-Science}, and non-equilibrium transport in QPCs has been observed to disrupt intriguing correlated electron transport \cite{Cronenwett-0.7}.  Despite the importance of understanding how non-equilibrium electrons decay energetically into a 2DEG, few experimental techniques can probe dynamics on the scale of the e-e scattering length $l_{e-e}$ directly.

Scanning gate microscopy (SGM) provides direct spatial information about both elastic and inelastic scattering of electrons \cite{Topinka-Science,Topinka-Nature,LeRoy-prism}.  The theory describing the e-e scattering rate \cite{Giuliani,Chaplik} has been validated by both SGM \cite{LeRoy-prism} and experiments using only lithographically-patterned gates \cite{Yacoby-DoubleSlit,Predel-EEHeating,Muller-EEBallistic,Schapers-EEBallistic}.  In this paper, we present SGM measurements of non-equilibrium electron transport through a QPC into a high mobility 2DEG.  We study a higher mobility 2DEG with higher injection energies (relative to the Fermi energy) than the previous SGM experiment investigating e-e scattering \cite{LeRoy-prism}.

At low injection energies ($\apprle \ 1 \ \mathrm{mV}$), we find that the injected electrons scatter by small-angles because of the confined phase space for e-e scattering in 2D compared to three dimensions, as has been found in previous patterned gate measurements \cite{Yanovsky-Angle}.  At higher injection energies, we observe a previously unreported reversal in sign of our SGM signal: the differential conductance through the system is \textit{increased} by moving the SGM tip into electron flow and thereby back-scattering electrons.  We propose that this effect is the result of scattering between electrons in the beam and a non-equilibrium distribution of electrons near the injection point.

Previous transport measurements showed effects of 2DEG heating when injecting high-energy electrons \cite{Predel-EEHeating}, but these previous measurements were not sensitive to the spatial profile of heating.  Our SGM technique shows evidence that the region of highly non-equilibrium 2DEG responsible for high e-e scattering rates is located within $\sim 1 \ \mathrm{\mu m}$ of the injection point, and we model the system with an effective temperature near the injection point.

\section{Experimental Procedure}

\begin{figure}
\begin{center}
\includegraphics[width=3.375in]{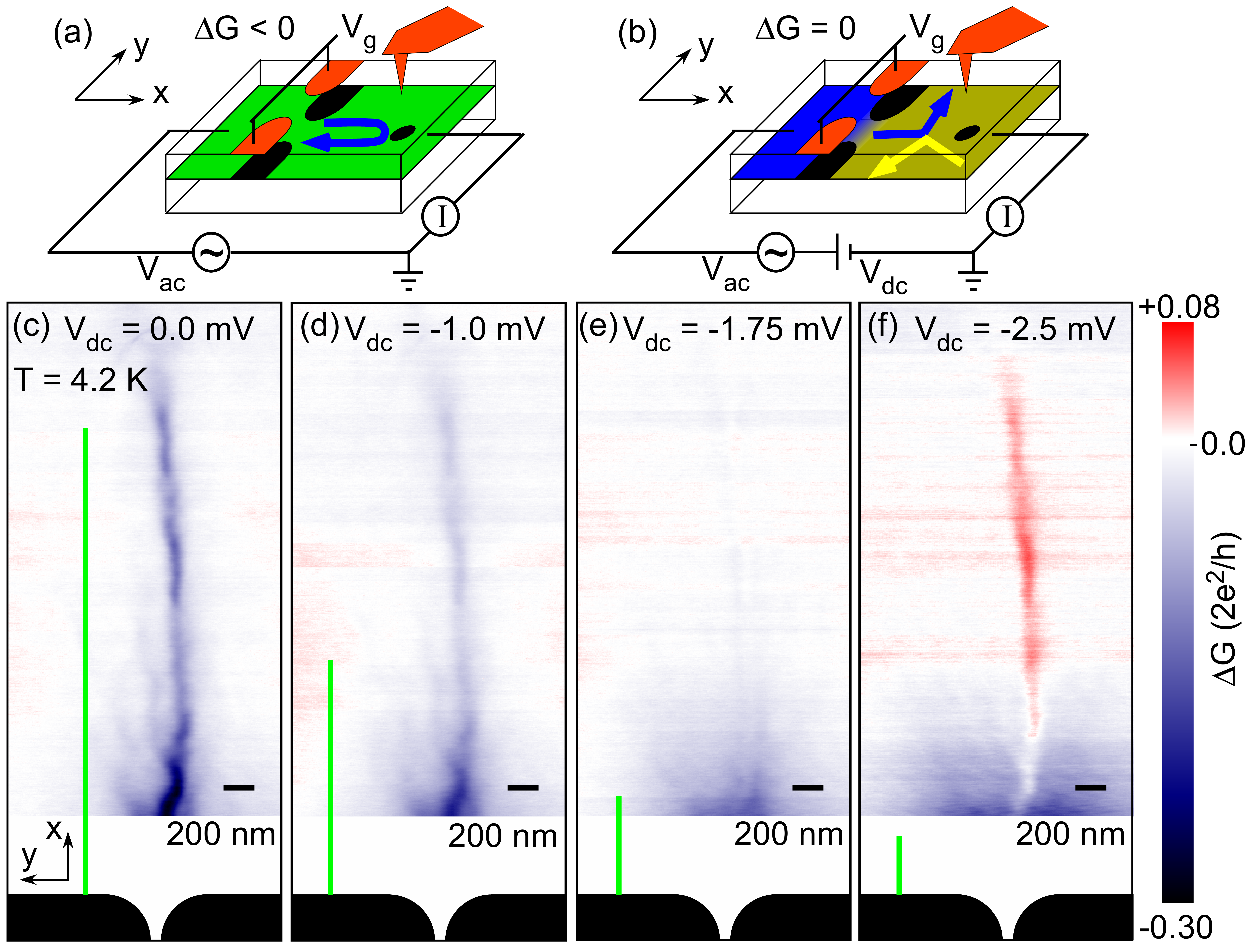}
\end{center}
\caption[Images of electron flow at varying injection energy]{Images of electron flow at varying injection energy.  (a) The standard imaging technique at $V_{dc} = 0$.  A QPC is defined in a 2DEG (green) with surface gates (orange), which deplete the 2DEG below (black regions).  A metallic SGM tip (orange) creates a depletion disk (black), which can scatter electrons emitted from the QPC back through the QPC (blue path), reducing the differential conductance: $\Delta G < 0$.  (b) The imaging technique to measure e-e scattering at high $|V_{dc}|$.  The injected electron (blue) can scatter with another electron in the 2DEG (yellow), and hence not complete a QPC-tip-QPC roundtrip.  With high e-e scattering, electrons are not backscattered through the QPC so $\Delta G = 0$.  (c)-(f) Images of electron flow all on the same color scale.  (c) Image of flow with minimal e-e scattering because $V_{dc} = 0$.  (f) At $V_{dc} = -2.5 \ \mathrm{mV}$, we observe the same flow pattern as in (c), but the signal has the opposite sign: $\Delta G > 0$.  This reversal in sign of SGM signal $\Delta G$ is unexpected and has not been reported previously.  The green bars in (c)-(f) are $l_{e-e}$ according to Eq. 1 ((d)-(f)) and Eq. 2 ((c)).}
\end{figure}

We use SGM to map electron flow emanating from a split-gate QPC \cite{Wharam,vanWees} into a 2DEG $100 \ \mathrm{nm}$ below the surface of a GaAs/AlGaAs heterostructure, with a bulk density of $1.5 \times 10^{11} \ \mathrm{cm}^{-2}$ and mobility of $4.4 \times 10^6 \ \mathrm{cm}^2/\mathrm{V s}$ at $4.2 \ \mathrm{K}$.  We employ standard SGM imaging techniques \cite{Topinka-Science,Topinka-Nature}, as depicted in Fig. 1(a).  We measure the differential conductance $G = dI/dV$ across the split-gate QPC using standard lock-in techniques.  Because the resistance through the system is dominated by the QPC, $G$ is determined by the transmission of electrons through the QPC.  The transmission or ``openness'' of the QPC is controlled by applying a voltage $V_g$ to the QPC gates.  We position a metallic SGM tip $\sim 30 \ \mathrm{nm}$ above the surface of the sample and apply a negative voltage to the tip, creating a depletion disk in the 2DEG below.  Previous SGM measurements of the Fermi wavelength in this device indicate that the 2DEG density is suppressed to $1.1 \times 10^{11} \ \mathrm{cm}^{-2}$ in the region between the tip and QPC because of the negatively charged tip and cantilever \cite{Jura-PRB}.

As we scan the SGM tip, we record the change in differential conductance $\Delta G = G - G_b$ where $G_b$ is the background conductance of the QPC in the absence of the tip.  When the SGM tip is above an area of high electron flow and there is minimal e-e (inelastic) scattering, the depletion disk scatters electrons back through the QPC and $G$ is reduced.  Thus, at zero bias $V_{dc} = 0$ across the QPC, spatial measurements of $\Delta G$ directly map electron flow with the correspondence that a negative $\Delta G(x,y)$ is proportional to the strength of current flow at $(x,y)$.  Images of electron flow at zero bias show branches because of elastic scattering off the disorder potential \cite{Topinka-Nature}.  In our sample, the disorder-determined elastic mean free path is $28 \ \mathrm{\mu m}$ but branches form over a much shorter length \cite{Jura-NaturePhysics}, as in Fig. 1(c) where one strong branch and a few weaker branches are visible.

We control e-e scattering by applying finite bias $V_{dc}$ across the QPC, which increases the phase space for e-e scattering.  For the injection energies we study (up to $3.5 \ \mathrm{meV}$, close to the Fermi energy of $3.9 \ \mathrm{meV}$), e-e scattering has been demonstrated to be the dominant inelastic scattering \cite{Yacoby-DoubleSlit,LeRoy-prism,Predel-EEHeating,Muller-EEBallistic,Schapers-EEBallistic}.  We neglect other mechanisms by which the injected electron can lose its excess energy, as discussed in Appendix A.  Electrons are injected with energy $\Delta = -|e| V_{dc}$ relative to the Fermi energy of the 2DEG, and the e-e scattering rate has been calculated to be \cite{Giuliani}:

\begin{equation}
\frac{1}{\tau_{e-e}} = \frac{E_F}{4 \pi \hbar} \left( \frac{\Delta}{E_F} \right)^2 \left[ \ln \left( \frac{E_F}{\Delta} \right) + \ln \left( \frac{2Q_{TF}}{k_F} \right) + \frac{1}{2} \right]
\end{equation}

\noindent where $k_F$ is the Fermi wavenumber, $Q_{TF} = 2 m e^2/\epsilon \hbar ^2 = 2\pi / 32 \ \mathrm{nm}$ is the 2D Thomas-Fermi screening wavenumber, $m$ is the effective mass, $\epsilon$ is the dielectric constant, $k_B T \ll \Delta \ll \hbar^2 k_F Q_{TF}/m$, and  $l_{e-e} = v \tau_{e-e}$ where $v$ is the electron velocity.

In the absence of e-e scattering, we would expect to use SGM to image how high-energy electrons injected into the disorder potential flow along branches.  We adjust the negative voltage on the tip so that it can still backscatter these high-energy electrons.  However, e-e scattering can alter an injected electron's path in a probabilistic manner, so that the injected electron no longer completes a QPC-tip-QPC roundtrip, restoring $G$ toward $G_b$ as fewer injected electrons complete the roundtrip (Fig. 1(b)).  Hence, for tip positions where $\Delta G < 0$ at $V_{dc} = 0$, we expect the SGM signal to fade to $\Delta G = 0$ as $|V_{dc}|$ and e-e scattering increase.  The SGM signal will decay away from the QPC over a distance $L_{e-e}$ that depends on $l_{e-e}$ (Eq.1). $L_{e-e} = l_{e-e}/2$ if a single e-e scattering event along the QPC-tip-QPC roundtrip path is sufficient to deflect an injected electron off the path.  Our measurement is sensitive to changes in the momentum direction, complementary to a previous SGM experiment studying e-e scattering, which relied on the loss of electron energy \cite{LeRoy-prism}.

\section{Experimental Data}

Figures 1(c)-(f) show images of electron flow at increasing $|V_{dc}|$ at $4.2 \ \mathrm{K}$.  The QPC gates are biased so that $G_b = 2e^2/h$.  The SGM signal $|\Delta G|$ is strongest at $V_{dc} = 0 \ \mathrm{mV}$  (Fig. 1(c)).  This observation is expected because e-e scattering is the lowest at zero bias where some e-e scattering still remains due to the finite temperature (see Eq. (2) below).  The SGM signal is reduced at $V_{dc} = -1.0 \ \mathrm{mV}$ (Fig. 1(d)), as expected because of increased e-e scattering.  However, the signal does not fade significantly beyond $\sim 1 \ \mathrm{\mu m}$ away from the QPC and persists out to a distance several times $l_{e-e}/2$ ($l_{e-e}$ denoted by green bars).  At $V_{dc} = -1.0 \ \mathrm{mV}$, the rate of deflection from the QPC-tip-QPC roundtrip path is thus not as fast as $1/\tau_{e-e}$ and a single e-e scattering event does not deflect an injected electron from the QPC-tip-QPC roundtrip path.  This observation is consistent with previous measurements showing that because of the confined phase space for e-e scattering in 2D, e-e scattering occurs predominantly at small angles $\theta \sim \sqrt{\Delta/E_F}$ \cite{Yanovsky-Angle}.  At $V_{dc} = -1.75 \ \mathrm{mV}$ (Fig. 1(e)), the SGM signal has nearly completely disappeared.  Increasing the bias further to $V_{dc} = -2.5 \ \mathrm{mV}$ (Fig. 1(f)), we observe a surprising effect: the flow pattern seen at $V_{dc} = 0 \ \mathrm{mV}$ reappears, but with opposite sign (i.e., $\Delta G > 0$).  We note that the bottom of Fig. 1(f) includes a slight, wide suppression of $\Delta G$ (blue), which is due to capacitive coupling from the tip negatively gating the QPC.  We correct for the spatial dependence of this capacitive tip-QPC coupling (see Ref. \cite{Jura-PRB} supplementary information for details), but at high bias the QPC is more sensitive to slight changes in gating and thus some of this effect is still visible.  This capacitive correction depends almost entirely on radial distance between the tip and QPC, and so cannot create features with angular spatial variations like the branches of current.

Measuring $\Delta G > 0$ means that the differential conductance is increased by moving the tip from a location with no electron flow into a location where there was electron flow at zero bias.  Usually the differential conductance $G$ is considered to be proportional to the transmission coefficient for electrons just at the electrochemical potentials of the two reservoirs on either side of a mesoscopic device \cite{Datta}.  It is therefore unexpected that introducing the tip into the electron flow and thus backscattering electrons should increase $G$.  As a point of clarification, we find only the \textit{differential} conductance $G = dI/dV$ increases when the tip is introduced into the current path; the total current $I$ or absolute conductance $I/V$ still decreases when the tip is introduced.  We observe $\Delta G > 0$ features at high $|V_{dc}|$ at either polarity, with the tip on either side of the QPC, and on other QPC devices in a different 2DEG.

\begin{figure}
\begin{center}
\includegraphics[width=3.375in]{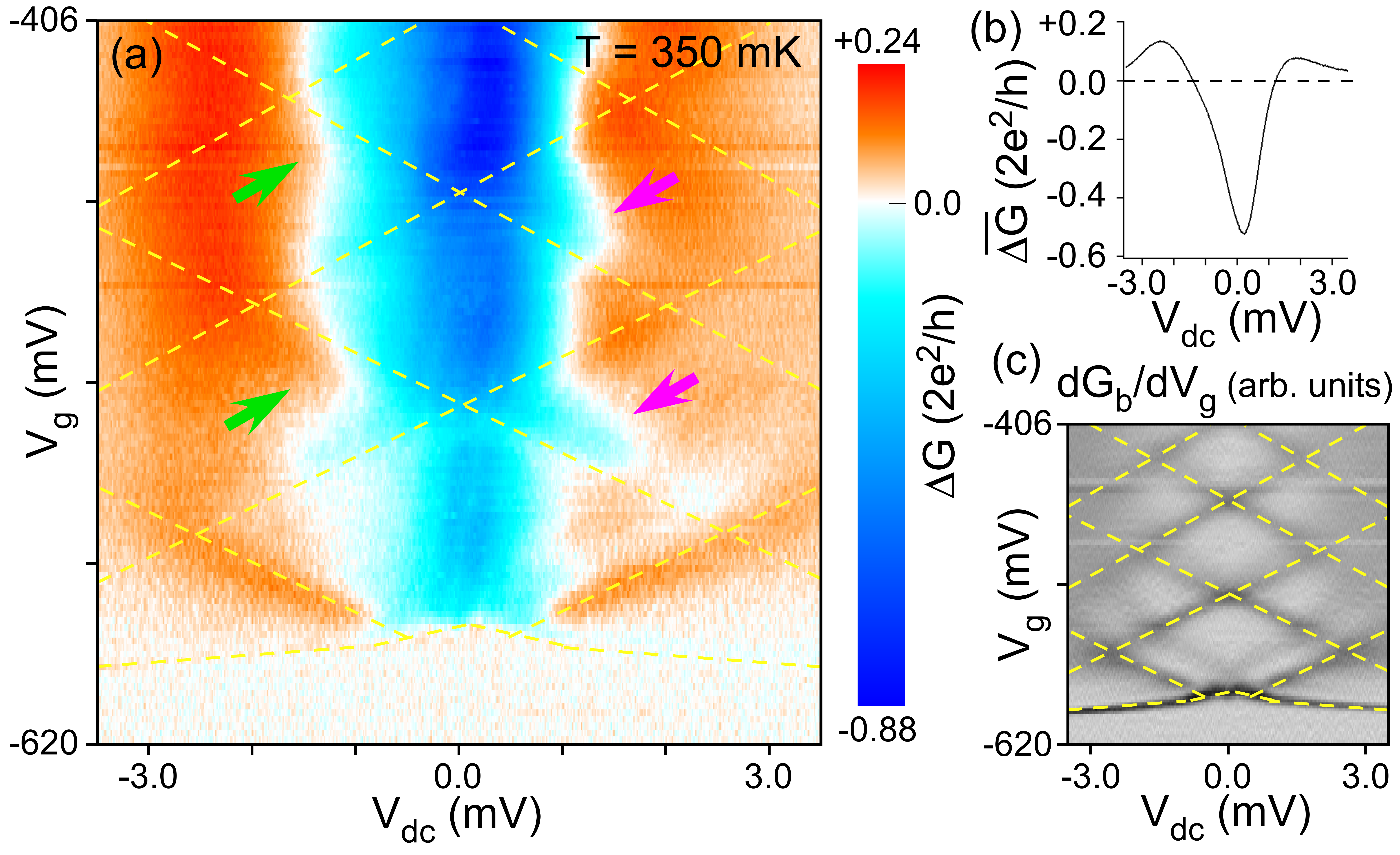}
\end{center}
\caption[Dependence of SGM signal on injection energy]{Dependence of SGM signal on injection energy.  (a) $\Delta G(V_{dc},V_g)$ at $L=1.0 \ \mathrm{\mu m}$ away from the QPC.  $\Delta G < 0$ for small $|V_{dc}|$, and $\Delta G > 0$ for large $|V_{dc}|$.  Green and purple arrows denote features associated with the state of the QPC (yellow dashed lines show transport through different sub-bands, as determined in (c)).  (b) $\overline{\Delta G}(V_{dc})$ averaged over $V_g$ in (a) from the bottom of the first sub-band to $-406 \ \mathrm{mV}$.  (c) $dG_b/dV_g$ of transport through the QPC without the tip blocking electron flow.  The yellow dashed lines denote the opening of transport through different sub-bands of the QPC.}
\end{figure}

To elucidate the origins of the enhancement $\Delta G > 0$, we measure $\Delta G$ at $350 \ \mathrm{mK}$ as a function of $V_{dc}$ and QPC sub-band occupation, controlled  by $V_g$.  We measure $G(V_{dc},V_g)$ with the tip in the electron flow and $G_b(V_{dc},V_g)$ with the tip out of the electron flow, subtracting the results to determine $\Delta G(V_{dc},V_g)$.  At each $V_g$, we measure $G(V_{dc})$, alternate the tip position, then measure $G_b(V_{dc})$, alternate the tip position back, and then step $V_g$.  Thus, any changes to the overall QPC conductance during the measurement do not change $\Delta G(V_{dc},V_g)$ after a single scan line at fixed $V_g$.  Reference \cite{Jura-PRB} shows more details of this measurement, including an example of the two tip locations, in and out of the electron flow.  In Fig. 2(a), we show $\Delta G(V_{dc},V_g)$ at $L=1.0 \ \mathrm{\mu m}$ away from the QPC.  The data are taken on a different thermal cycle from room-temperature down to $350 \ mK$ than those in Fig. 1.

For $|V_{dc}|$ near $0 \ \mathrm{mV}$, $\Delta G$ is negative, but for $|V_{dc}| \apprge 1.5 \ \mathrm{mV}$, $\Delta G$ is positive.  $\Delta G$ mostly depends on $V_{dc}$, but there is some dependence on $V_g$ (i.e., the openness of the QPC).  Figure 2(b) shows $\overline{\Delta G} (V_{dc})$, averaged over $V_g$ in Fig. 2(a) to remove effects of specific geometries.  The asymmetry of $\overline{\Delta G} (V_{dc})$ near $V_{dc} = 0 \ \mathrm{mV}$ may be caused by interference effects between the tip and QPC, which, because of the particular tip location, tend to emphasize negative $\overline{\Delta G} (V_{dc})$ at slightly positive $V_{dc}$ \cite{Jura-PRB}.  Other measurements of $\overline{\Delta G} (V_{dc})$ (see Fig. 3) display less asymmetry.  In Fig. 2(c), we plot $dG_b/dV_g$, showing diamond-like features (denoted with yellow dashed lines) which are a result of transport through different sub-bands of the QPC \cite{Kouwenhoven-QPCDiamond} (see Appendix B for more characterization of the QPC).  We place yellow dashed lines at the same $(V_{dc},V_g)$ locations in Fig. 2(a).  Purple arrows denote locations where $\Delta G < 0$ (blue) features jut out above yellow dashed lines, and green arrows denote locations where $\Delta G > 0$ (red) features jut in below yellow dashed lines.

\begin{figure}
\begin{center}
\includegraphics[width=3.375in]{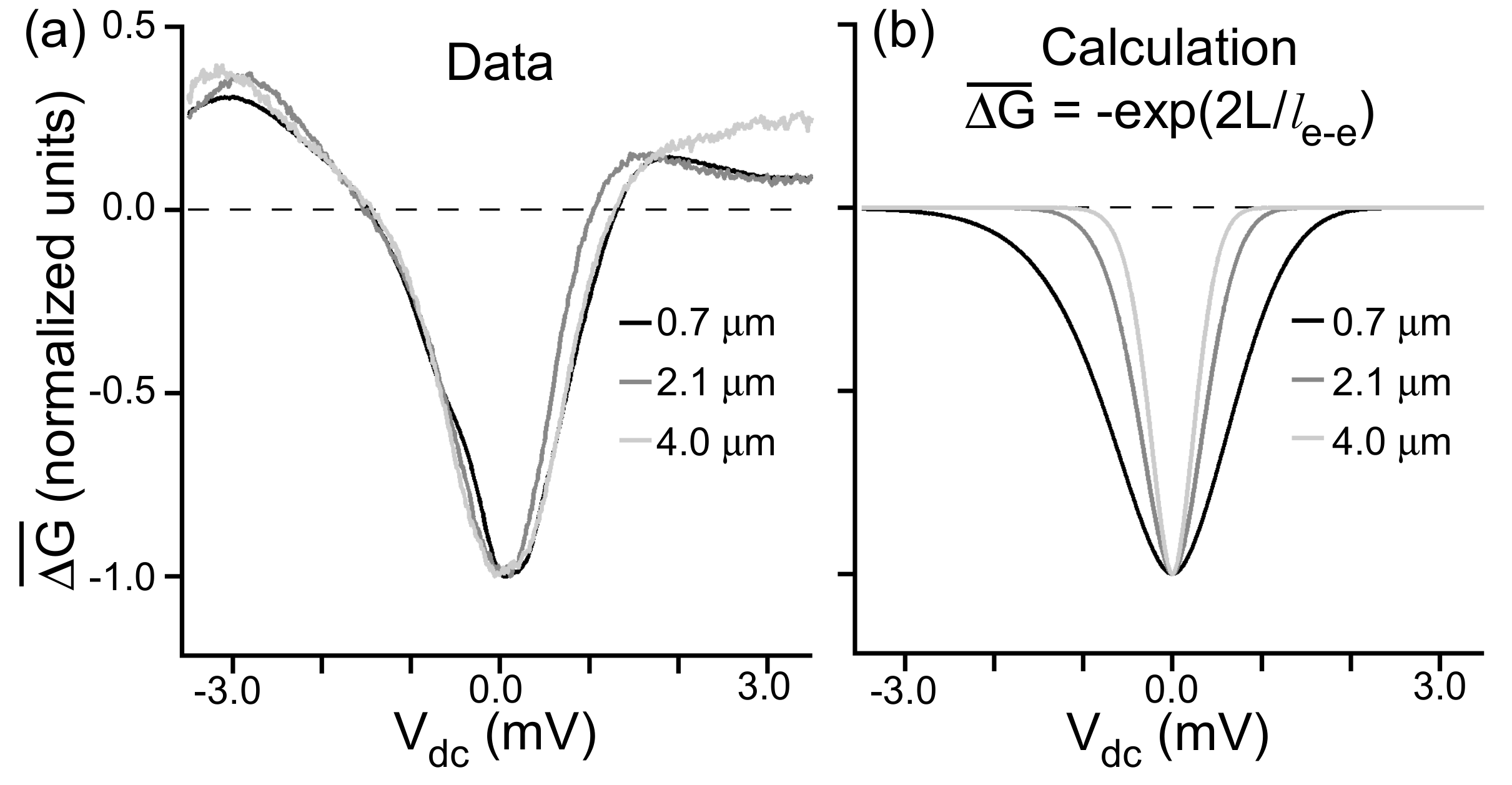}
\end{center}
\caption[Distance dependence of $\overline{\Delta G} (V_{dc})$]{Distance dependence of $\overline{\Delta G} (V_{dc})$.  (a) $\overline{\Delta G} (V_{dc})$ measured at various distances of $L$: $0.70 \ \mathrm{\mu m}$, $2.1 \ \mathrm{\mu m}$, and $4.0 \ \mathrm{\mu m}$.  The curves are normalized to the minimum value and all lie on top of each other well.  Therefore, all e-e scattering dictating the shapes of these curves occurs within $0.70 \ \mathrm{\mu m}$ of the QPC.  (b) Calculated $\overline{\Delta G} (V_{dc})$ assuming e-e scattering according to Eq. 1 and a single e-e scattering event anywhere on the QPC-tip-QPC roundtrip deflects the injected electron off the path.  The curves have different widths at different distances and do not reproduce the $\overline{\Delta G} > 0$ observed in the data.}
\end{figure}

In Fig. 3 we investigate $\overline{\Delta G} (V_{dc})$, measured the same way as in Fig. 2, at various distances $L$ away from the QPC on two different temperature cycles from room-temperature to $350 \ mK$.  Because increasing $L$ increases the probability that an electron scatters along the QPC-tip-QPC roundtrip, we expect $|\overline{\Delta G} (V_{dc})/\overline{\Delta G} (0 \ mV)|$ to be smaller for longer $L$.  Figure 3(a) shows $\overline{\Delta G} (V_{dc})$ measured at $L = 0.70 \ \mathrm{\mu m}$, $2.1 \ \mathrm{\mu m}$, and $4.0 \ \mathrm{\mu m}$.  The $\overline{\Delta G}$ curves are normalized to have the same minimum value to account for the strength of flow at different locations.  There is no offset so that $\overline{\Delta G} = 0$ still has the same meaning, and $V_{dc}$ is not scaled.  Strikingly, the scaled $\overline{\Delta G} (V_{dc})$ curves have nearly identical widths and $\overline{\Delta G} > 0$ features at all distances.  The lack of distance dependence suggests that the e-e scattering responsible for the observed $\overline{\Delta G}$ features occurs within $0.70 \ \mathrm{\mu m}$ of the QPC.

\section{Failed Simple Model}

A simple common treatment of e-e scattering that only considers electrons injected at $\Delta = -|e|V_{dc}$ does not reproduce the bias dependence.  In this picture, a single e-e scattering event anywhere on the QPC-tip-QPC roundtrip path deflects the injected electron, so $\overline{\Delta G} \propto -\exp(-2L/l_{e-e})$ where $l_{e-e}$ is calculated according to Eq. 1.  The calculated averaged conductance curves in Fig. 3(b) using this equation do not reproduce the $\overline{\Delta G} > 0$ features, nor the lack of distance dependence in the widths of curves.

The widths of $\overline{\Delta G} (V_{dc})$ curves are related to the rates at which injected electrons are deflected off the QPC-tip-QPC roundtrip, and the data in Fig. 3(a) are evidence for small-angle e-e scattering.  Wider $\overline{\Delta G}(V_{dc})$ curves suggest lower rates of deflection off the QPC-tip-QPC roundtrip.  The rate of deflection off the QPC-tip-QPC roundtrip is related to the e-e scattering rate, but they are not necessarily the same.  If a single e-e scattering event along the QPC-tip-QPC roundtrip deflects an injected electron off this path, we expect a decay $\overline{\Delta G} \propto -\exp(-2L/l_{e-e})$ where the factor of $2$ is from the roundtrip path.  At the farthest distance, $L = 4.0 \ \mathrm{\mu m}$ (Fig. 3(a), light gray curve), our data have a full-width at half-minimum (FWHM; i.e., the width at $\overline{\Delta G} = -0.5$ in normalized units) of $1.41 \ \mathrm{mV}$.  However, the FWHM in the calculation $\overline{\Delta G} \propto -\exp(-2L/l_{e-e})$ (Fig. 3(b), light gray curve) is $0.58 \ \mathrm{mV}$, significantly narrower than our data, indicating a single e-e scattering event along the QPC-tip-QPC roundtrip does not cause a loss of SGM signal $\Delta G$.

For another calculation, we can instead assume conditions that would produce a wider FWHM: the electron density in the imaging region is that of the bulk, and electrons reflect off the tip as spherical waves, with the result that an e-e scattering event on the path from the tip back to the QPC may not significantly change $\overline{\Delta G}$.  Under this latter assumption, only an e-e scattering event on the outgoing path from the QPC to tip will deflect injected electrons off the QPC-tip-QPC roundtrip, and we expect $\overline{\Delta G} \propto -\exp(-L/l_{e-e})$.  Under these wider FWHM conditions, we calculate a FWHM of $1.11 \ \mathrm{mV}$ for $L = 4.0 \ \mathrm{\mu m}$, still narrower than our data.  Thus to explain the observed $\overline{\Delta G}(V_{dc})$ at longer distances from the QPC, we must invoke small-angle e-e scattering: an injected electron can experience e-e scattering but only be deviated by a small angle and still complete the QPC-tip-QPC roundtrip.  Furthermore, e-e scattering has been observed to occur at the small angles $\theta \sim \sqrt{\Delta/E_F}$ in 2D \cite{Yanovsky-Angle}, and so small-angle scattering becomes more important at lower injection energies $\Delta$, the energies at which we expect to detect e-e scattering only at longer distances.

\section{Mechanism for Enhancement of Differential Conductance with the Tip}

\begin{figure}
\begin{center}
\includegraphics[width=3.375in]{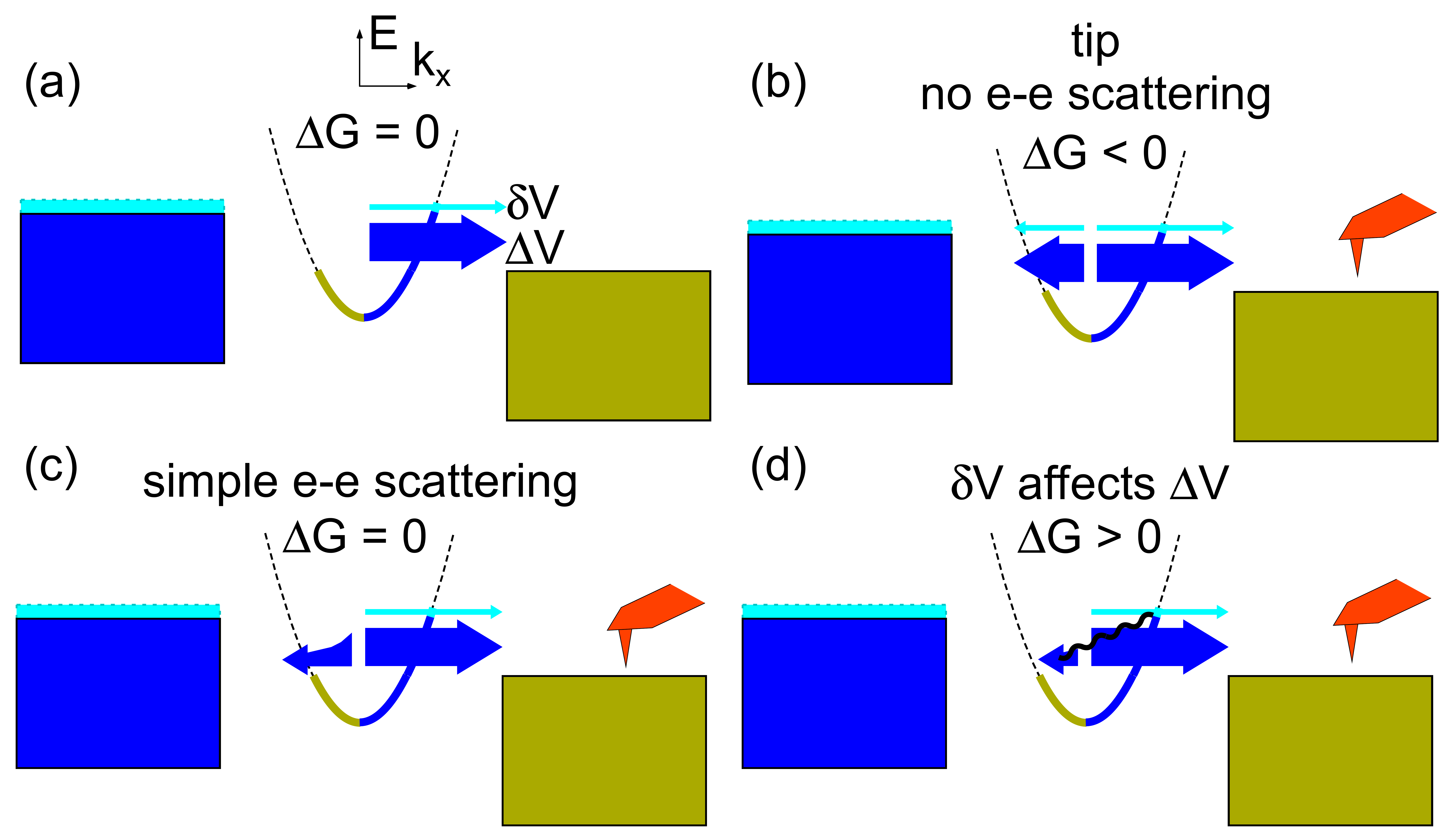}
\end{center}
\caption[General mechanism for $\Delta G > 0$]{General mechanism for $\Delta G > 0$.  Energy space pictures for transport through the QPC.  $\Delta G$ is proportional to the amount of net electron current flow through the QPC in response to and including $\delta V$ electron current, minus the amount of $\delta V$ electron current with no tip present (light blue arrow in (a)).  (a) The energy picture before e-e scattering or scattering from the tip are considered.  There are $\delta V$ and $\Delta V$ electron currents in the $+k_x$ direction, and $\Delta G = 0$ because there is no tip present.  (b) Now scattering from the tip is considered.  A fraction of the current is scattered back through the QPC in the $-k_x$ direction.  The net light blue arrow is not as long as that in (a), so $\Delta G < 0$.  (c) Now a simple picture of e-e scattering, equivalent to that in Fig. 3(b), is considered: high energy $\delta V$ and $\Delta V$ electrons are more likely to experience e-e scattering with the right 2DEG reservoir into which they are injected, and therefore less likely to complete the QPC-tip-QPC roundtrip.  There is less electron current in the $-k_x$ direction at higher energies, and in the $\delta V$ energy window, there is no $-k_x$ electron current.  The net light blue arrow is the same length as in (a), so $\Delta G = 0$.  (d) Now a more complete picture of e-e scattering is considered: the addition of $\delta V$ electrons makes it less likely for $\Delta V$ electrons to complete the QPC-tip-QPC roundtrip.  There are fewer $\Delta V$ electrons moving in the $-k_x$ direction through the QPC.  Thus, the net electron current through the QPC in response to the addition of $\delta V$ electrons includes more than just the $\delta V$ electrons; the net electron current also includes the lack of $\Delta V$ electron current in the $-k_x$ direction.  Thus, $\Delta G > 0$.}
\end{figure}

The previous simple calculation fails to explain $\Delta G > 0$ features because it does not account for how \textit{all electrons} injected in the energy window between $0$ and $\Delta$ scatter.  We next present a general mechanism by which $\Delta G$ features can arise.  We designate ``$\Delta V$'' electrons as those driven by the steady voltage $V_{dc}$ in the energy window $\Delta$.  We designate ``$\delta V$'' electrons as those driven by $V_{ac}$ within a small band of energy close to $\Delta$.  Though it is often assumed that $G$ is determined by the transport of $\delta V$ electrons, $\delta V$ electrons can affect the transmission of $\Delta V$ electrons through the QPC as well.  $\Delta G > 0$ arises if the addition of $\delta V$ electrons causes a responding net current flow through the QPC of more than just $\delta V$ electrons alone.  With the tip in the path of electrons, this is possible because low energy $\Delta V$ electrons are reflected back through the QPC.  As shown in Fig. 4, if $\delta V$ electrons do not complete the QPC-tip-QPC roundtrip and also prevent some $\Delta V$ electrons from completing the roundtrip, $\Delta G$ will be greater than $0$.

Figure 4 shows a schematic for transport across the QPC.  In all diagrams, the left 2DEG reservoir is represented with blue, and the right 2DEG reservoir with green, corresponding to the colors in Figs. 1(a) and 1(b).  The vertical axis is energy, and the horizontal axis is a combination of real and momentum space.  Quasi-one-dimensional electronic states through the QPC are represented with the dashed parabola, indicating the dispersion relation $E \propto k_x^2$.  The electrochemical potential of the left reservoir is $-|e|(V_{dc} + V_{ac})$ above the electrochemical potential of the right reservoir, where $V_{ac}$ is the small time-oscillating voltage used in the lock-in measurement.  The difference in electrochemical potentials between the reservoirs drives an electron current from left to right through the QPC.  Electron currents are designated with arrows; the width of an arrow indicates electron flow in that energy window, and the length of the arrow is proportional to the amount of electron flow per unit energy.  $\Delta V$ electrons, driven by $V_{dc}$, are designated with a dark blue arrow, and $\delta V$ electrons, driven by the small $V_{ac}$, are designated with a light blue arrow.  The caption of Fig. 4 explains how $\Delta G >0$ can arise if $\delta V$ electrons affect the transmission of $\Delta V$ electrons.

\begin{figure}
\begin{center}
\includegraphics[width=3.375in]{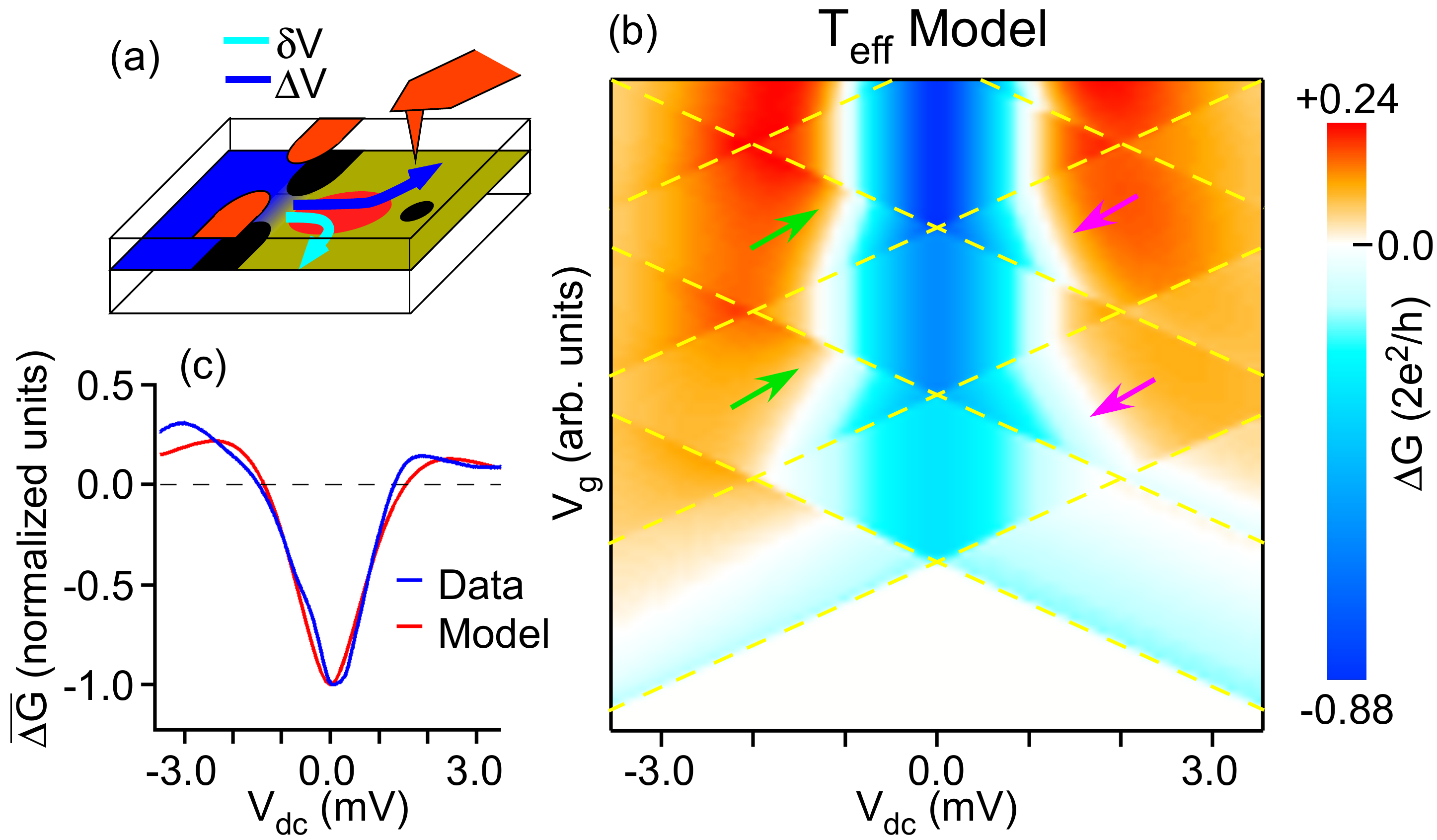}
\end{center}
\caption[Effective temperature model calculations]{Effective temperature model calculations.  (a) Schematic origin of $\Delta G > 0$ features.  Injected electrons create a region of 2DEG near the QPC with a highly non-equilibrium electron distribution (red).  We approximate this region with an effective temperature $T_\mathrm{eff}$ that is significantly hotter than surroundings.  The injection of high-energy $\delta V$ electrons causes an increase in $T_\mathrm{eff}$, which then scatters lower energy $\Delta V$ electrons off the QPC-tip-QPC roundtrip.  (b) $\Delta G(V_{dc},V_g)$ according to the $T_\mathrm{eff}$ model for a region of ``hot'' 2DEG extending away from the QPC by $L_T = 0.7 \ \mathrm{\mu m}$.  The model shows features similar to those in the data of Fig. 2(a).  The purple and green arrows denote features associated with the yellow diamonds, as in Fig. 2(a).  Crossing a yellow line indicates the opening of a new QPC sub-band and thereby a change in the ratio of $\delta V$ to $\Delta V$ electrons.  (c) $\overline{\Delta G} (V_{dc})$ of data (from Fig. 3(a) at $L = 0.7 \ \mathrm{\mu m})$) and model in panel (b), averaged over the same ranges of $V_g$.  The model reproduces $\Delta G > 0$ features with approximately the correct magnitude and $V_{dc}$-dependence.}
\end{figure}

We next describe a mechanism by which $\delta V$ electrons can affect the flow of $\Delta V$ electrons between the QPC and tip.  Transport at finite bias creates a non-equilibrium region in the 2DEG near the QPC as injected electrons scatter with the 2DEG.  This non-equilibrium region of 2DEG has increased phase space available for e-e scattering of other injected electrons.  The injection of $\delta V$ electrons drives this region of 2DEG further out of equilibrium and enhances the scattering of $\Delta V$ electrons, as depicted in Fig. 5(a).  Measurements of non-equilibrium transport through QPCs have shown that regions of 2DEG several micrometers away can be heated by several kelvin by the out-of-equilibrium electrons \cite{Predel-EEHeating}.  Increased temperature leads to increased e-e scattering rates, as reflected in Eq. 2.  In order to better understand our data, we approximate the region of 2DEG with partially occupied states near $E_F$ as having a higher effective temperature $T_\mathrm{eff}$ over a distance $L_T$ away from the QPC, where we know experimentally $L_T \apprle 0.7 \ \mathrm{\mu m}$.  This $T_\mathrm{eff}$ model neglects the expected anisotropy of the electron momentum distribution as it evolves out of the QPC and simply assumes excess energy is converted to an effective temperature.  An effective temperature model has the advantage that e-e scattering rates have previously been calculated for a thermal distribution of electrons \cite{DasSarma-Tee}.  By fitting the effective temperature, we reproduce many of the features of our experimental data showing $\Delta G > 0$, and we learn the extent to which this region of 2DEG near the QPC is driven out of equilibrium.

The e-e scattering rate caused by a thermal distribution has been calculated \cite{DasSarma-Tee} to explain tunneling measurements \cite{Murphy} as:

\begin{equation}
\frac{1}{\tau_{e-e}} = \frac{\pi E_F}{8 \hbar} \left( \frac{k_B T}{E_F} \right)^2 \ln \left( \frac{E_F}{k_B T} \right)
\end{equation}

\noindent for $\Delta << k_B T << E_F$.

In our $T_\mathrm{eff}$ model, we parametrize the dependence of effective temperature on bias voltage as $T_\mathrm{eff} = \alpha V$ for a single sub-band of conductance (multiple sub-bands add temperature in quadrature).  We calculate the transmission of all electrons with energy $\xi$ in the range $0 < \xi < \Delta$ as a function of $V_{dc}$ and $V_g$, using a constant sub-band spacing of $2 \ \mathrm{meV}$.  We assume the transmission of each electron is $\exp(-2L/l_{e-e})$.  Unless noted otherwise, we calculate $l_{e-e}$ according to Eq. 2 with $T = T_\mathrm{eff}$.  We previously showed that the rate of deflection off the QPC-tip-QPC roundtrip is slower than the rate in Eq. 1 at low injection energies ($\apprle 1 \ \mathrm{meV}$), which were more relevant at longer distances ($\apprge 4 \ \mathrm{\mu m}$).  Here we are concerned with relatively larger injection energies at closer distances, so we also assume that a single e-e scattering event on the roundtrip QPC-tip-QPC deflects the injected electron off this path (i.e. that the transmission of each electron is $\exp(-2L/l_{e-e})$).  The physical motivation for the effective temperature model is that $k_B T_\mathrm{eff}$ comes from the injection energy $\Delta$, which suggests for consistency $k_B T_\mathrm{eff} < \Delta$.  For electrons in the energy range $0 < \xi < k_B T_\mathrm{eff}$, Eq. 2 is valid.  For higher energy electrons $k_B T_\mathrm{eff} \apprle \xi < \Delta$, the inequality required for Eq. 2 is contradicted because Eq. 1 is faster (shorter $l_{e-e}$).  In a first calculation we determine $l_{e-e}$ by Eq. 2 only and recognize it may be an underestimate for the e-e scattering rate for the electrons at the highest injection energies; in a later calculation, we determine $l_{e-e}$ for an electron with energy $\xi$ using whichever is the higher scattering rate of Eq. 1 or 2 for that injection energy.  We fit $\alpha$ and never $l_{e-e}$.

In Figs. 5(b) and 5(c), we display the calculated $\Delta G$ according to this $T_\mathrm{eff}$ model.  The simple $T_\mathrm{eff}$ model in Figs. 5(b) and 5(c) shows agreement with our data, including $\Delta G > 0$ at high $|V_{dc}|$ and features (denoted with green and purple arrows) that are associated with the sub-band opening (yellow dashed lines).  In this $T_\mathrm{eff}$ model, $\Delta G = 0$ when $V_{dc}$ increases the temperature enough so that as many back-reflected $\Delta V$ electrons are prevented from completing the QPC-tip-QPC roundtrip as $\delta V$ electrons are still able to complete the roundtrip.  The asymmetry in the simulation comes from the different velocity of electrons at negative versus positive bias, as well as whether the injection electrochemical potential is above (negative bias) versus below (positive bias) the bottom of the parabolic dispersion relation of sub-bands at half-plateaus.

We fit $\alpha = 2.6 \ \mathrm{K}/\mathrm{mV}$ for $L_T = 0.7 \ \mathrm{\mu m}$.  In fitting $\alpha$ we assume a constant effective temperature over the distance $L_T$, but the actual electron distribution should be smoothly varying with position.  We do not have data for the spatial details of the effective temperature profile for $L < 0.7 \ \mathrm{\mu m}$, but we define a different parameter $\beta$ which contains information about $\int \left[T(x)\right]^2 dx$, the term that scales as the total amount of scattering, neglecting small logarithmic variations from the $\ln \left( \frac{E_F}{k_B T} \right)$ factor.  We therefore define $\beta = \alpha \sqrt{L_T}$, giving $\beta = 2.4 \ \mathrm{K} \sqrt{\mathrm{\mu m}}/\mathrm{mV}$.  If we instead assume that an electron with energy $\xi$ scatters according to the faster of Eq. 1 or Eq. 2, as discussed previously, we fit $\alpha = 1.8 \ \mathrm{K}/\mathrm{mV}$ for $L_T = L = 0.7 \ \mathrm{\mu m}$.  Previous transport measurements found a temperature increase of roughly $1.2 \ \mathrm{K}/\mathrm{mV}$ but over a distance of $3.4 \ \mathrm{\mu m}$ away from a QPC \cite{Predel-EEHeating}.

The effective temperature model does not take into account the anisotropic momentum distribution of electrons in the 2DEG immediately after the injection point, and we emphasize that we only fit the dependence on bias voltage of an \textit{effective temperature} needed to explain e-e scattering.  More realistically, we expect the momentum distribution to be deformed so that there are more electrons traveling in the $+k_x$ direction and fewer in the $-k_x$ direction.  This filling of states implies the 2DEG will more efficiently scatter electrons moving in the $-k_x$ direction (electrons returning from the tip towards the QPC).  While our $T_\mathrm{eff}$ model assumes an e-e scattering rate that is equal for both ``outbound'' electrons (along the QPC-tip path) and ``inbound'' electrons (along the tip-QPC path), we thus expect the scattering rate for ``inbound'' electrons to be higher in a more realistically detailed model.  We also recognize that strong scattering may occur very close to or inside the QPC where interactions between sub-bands may play a role.

\section{Conclusions}

We have measured the spatial extent of the non-equilibrium distribution resulting from a beam of high-energy electrons injected into a 2DEG, and we have observed surprising conductance behavior associated with non-equilibrium transport.  We reproduce the main features of our data using a simple $T_\mathrm{eff}$ model and an assumed isotropic distribution of electron momenta.  Further theoretical work and experiments that discriminate between outgoing and incoming electrons may increase our understanding of non-equilibrium phenomena in QPCs and 2DEGs.

\section{Acknowledgments}

We thank A. Sciambi for help characterizing the 2DEG sample.  We are grateful to Y. Oreg, D. Meidan, D. Loss and S. Kivelson for useful discussions.  This work was funded by the Center for Probing the Nanoscale (CPN), an NSF NSEC, NSF Grant Nos. PHY-0830228 and PHY-0425897.  Work was carried out in part at the Stanford Nanofabrication Facility of NNIN supported by NSF, grant ECS-9731293.  M.P.J. acknowledges support from the NDSEG program.  D.G.-G. recognizes support from the David and Lucile Packard Foundation.\newline

\noindent $^\ast$ Present address: Bandgap Engineering, Woburn, MA 01801

\noindent $^\dag$ Present address: Hitachi GST, San Jose, CA 95135

\noindent $^\ddag$ Present address: Alion Inc., Richmond, CA 94804

\noindent $^\S$ Present address: Department of Electrical Engineering, Princeton University, Princeton, New Jersey 08544

\noindent $^\parallel$ Corresponding author; goldhaber-gordon@stanford.edu 

\appendix
\section{Appendix A: Dominance of Electron-Electron Scattering}

When considering how the injected high-energy electrons inelastically scatter and lose energy, for the range of injection energies we study (up to $\Delta = 3.5 \ \mathrm{meV}$) we can neglect other sources of scattering, including plasmon emission and scattering with phonons.  The threshold energy $\Delta_C$ above which plasmon emission occurs was calculated to be (Eq. 17 in Ref. \cite{Giuliani}; see corrections in Ref. \cite{Muller-EEBallistic}):

\begin{eqnarray*}
\Delta_C = & \left[ \frac{16 \sqrt{2m} e^2 E_F \left( E_F + \Delta_C \right)^{1/2}}{3 \epsilon \hbar} \right] ^{1/2} \\
& \times \cos \left[ \frac{\pi}{3} + \frac{1}{3} \arccos \left( \frac{E_F + \tilde{\Delta}}{E_F + \Delta_C} \right) ^{3/4} \right]
\end{eqnarray*}

\noindent where $\tilde{\Delta} = \left[ \left( \frac{27 r_S}{8 \sqrt{2}} \right) ^{2/3} - 1 \right] E_F$.  $r_S$ is the ratio of the interelectronic distance $1/\sqrt{\pi n}$ to the effective Bohr radius $a_0 \epsilon m_0/m = 10.3 \ \mathrm{nm}$.  We find $\Delta_C = 1.7 \ E_F = 6.7 \ \mathrm{me}V$, larger than the highest injection energy we use.  The threshold injection energy above which longitudinal optical (LO) phonons are emitted is even larger: $36 \ \mathrm{meV}$ \cite{Sivan-LOphonon,Dzurak}.  For acoustic phonons, $l_{e-ph}$ is typically estimated to be on the order of $100 \ \mathrm{\mu m}$ for $\Delta \approx 1 \ \mathrm{meV}$ \cite{Predel-EEHeating,Schinner}.  If we assume the electron-phonon scattering rate scales as $1/\tau_{e-ph} \sim \Delta^3$ \cite{Mittal}, for $\Delta = 3.5 \ \mathrm{meV}$, $1/\tau_{e-e} >> 1/\tau_{e-ph}$ by an order of magnitude.

\section{Appendix B: QPC Characterization}

\begin{figure}
\begin{center}
\includegraphics[width=3.375in]{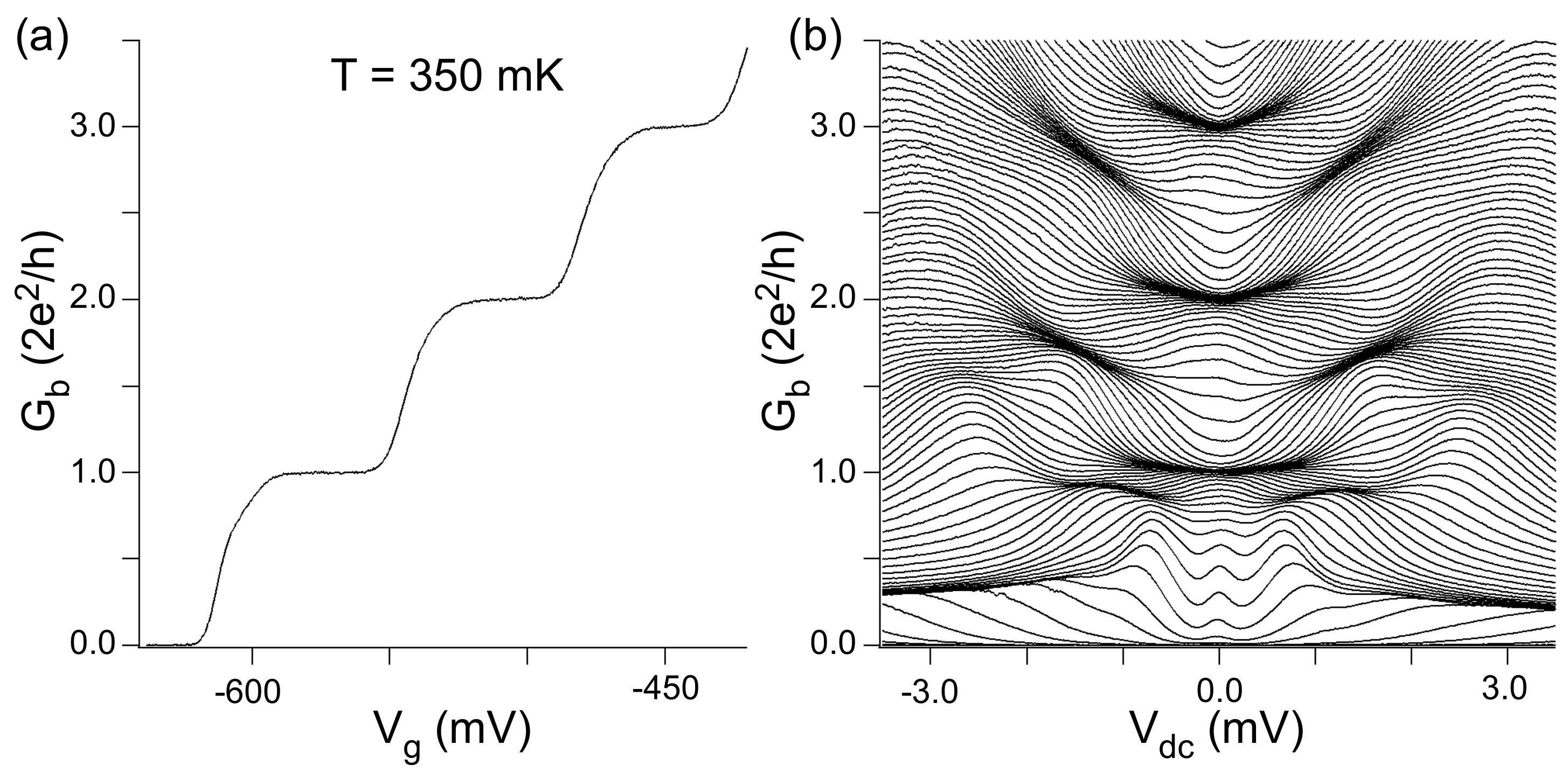}
\end{center}
\caption[Conductance data of QPC without tip]{Conductance data of QPC without tip.  (a) Conductance plateaus as a function of $V_g$, which controls the number of transmitting sub-bands.  (b) $G_b(V_{dc})$ for small steps in $V_g$ (i.e. $G_b(V_{dc},V_g)$) for the $L = 4.0 \ \mathrm{\mu m}$ measurement in Fig. 3(a).  The data have been corrected for series resistance leading to the QPC.}
\end{figure}

Here we present measurements of $G_b$, the conductance of the QPC alone with no tip blocking electron flow, in order to show that the electrical behavior is similar to that of other previously studied QPC devices.  Figure 6(a) shows the well-known conductance plateau of $G_b(V_g)$ \cite{Wharam,vanWees}, and Fig. 6(b), measured at small increments of the gate voltage $V_g$, shows $G_b(V_{dc})$ which displays features similar to those in Ref. \cite{Cronenwett-0.7}.

\newcommand{\noopsort}[1]{} \newcommand{\printfirst}[2]{#1}
  \newcommand{\singleletter}[1]{#1} \newcommand{\switchargs}[2]{#2#1}

\end{document}